# Dynamical Behaviors of Small-scale Buoyant Diffusion Flame Oscillators in Externally Swirling Flows

Tao Yang[1], Yuan Ma[1], Peng Zhang[2,*]

*1. Department of Mechanical Engineering, The Hong Kong Polytechnic University, Hung Hom, Kowloon, Hong Kong*
*2. Department of Mechanical Engineering, City University of Hong Kong, Kowloon Tong, Kowloon, Hong Kong*

## Abstract

Small-scale flickering buoyant diffusion flames in externally swirling flows were computationally investigated with a particular interest in identifying and characterizing various distinct dynamical behaviors of the flame oscillators under different swirling flow conditions. By varying the external swirl, six distinct flame dynamical modes, such as the flickering flame, the oscillating flame, the steady flame, the lifted flame, the spiral flame, and the flame with a vortex bubble, were computationally identified in both physical and phase spaces and analyzed from the perspective of vortex dynamics. Specifically, the frequency of buoyancy-induced flame flicker nonlinearly increases with the swirling intensity in the weak swirl regime. Further increasing the swirling intensity causes the vortex shedding to occur either around the flame tip or downstream of the flame, and the flame stops flickering but oscillates with small amplitude or stays in a steady state. A sufficiently high swirling intensity locally extinguishes the flame at its base, leading to a lifted flame. In addition, the spiral mode and the vortex-bubble mode were found for the flame at large swirl angles. Through establishing the phase portrait for featuring the flow in flames, the dynamical behaviors are presented and compared in phase space.

## Keywords

* Corresponding author
  E-mail address: penzhang@cityu.edu.hk
  Tel: (852)34429561

# 1 Introduction

The study of diffusion flames is closely related to the flame stability and fire safety of many industrial and environmental applications. The "flickering" or "puffing" of a buoyant diffusion flame has been of long-lasting research interest for several decades. The vibratory motion of Bunsen diffusion flames was first discovered and referred to as "the flicker of luminous flames" by Chamberlin and Rose [1]. A similar phenomenon in a Burke-Schumann diffusion flame [2] was described as "the vibration is seen to consist of a progressive necking of the flame which can lead to the formation of a flame bubble, which burns itself out separated from the anchored flame".

Previous studies [3-7] found that the flickering of flame is a buoyance-dominated flow phenomenon that is relatively insensitive to other flame parameters. As a result, the flicker frequency $f_0$ is proportional to $(g/D)^{1/2}$, where $g$ is the gravitational constant and $D$ the fuel inlet diameter, and the proportionality factor slightly varies among many fuels [4]. This correlation led to the dimensionless scaling law, $St \sim Fr^{-1/2}$, where $St = f_0 D/U_0$ is Strouhal number and $Fr = U_0^2/gD$ is the Froude number. Chen et al.'s flow visualization of a methane jet diffusion flame [8] provided a piece of confirmative experimental evidence to the vortex-dynamical physical picture that the luminous flame was elongated vertically and contracted horizontally by the outside large toroidal vortices (the buoyance-induced Kelvin-Helmholtz instability), causing the formation of a "neck" or even the pinch-off of the flame top. By directly calculating the dimensionless circulation $\Gamma$ of a toroidal vortex at the end of a flicker period $\tau$, Xia and Zhang [9] obtained

$$\Gamma(\tau) = C_h Ri\, St^{-2} + C_j Fr^{1/2} St^{-1} \qquad (1)$$

where $Ri = (\rho_\infty/\rho - 1)gD/V^2$ is Richardson number; $\rho$ and $\rho_\infty$ are the density of flame and ambient, respectively; $C_h$ and $C_j$ are constants for the advective speed of the vortex and the circulation addition by the inflow, respectively. By applying a vortex shedding criterion, $\Gamma(\tau) = C$, where $C$ is a system-dependent constant [10], they obtained a scaling law that generalizes the previous scaling laws and well predicts experimental data in the literature for $Fr \ll 1$ and $Ri \gg 1$. Based on the vortex-dynamical understanding of flickering flames, Yang and Zhang [11] recently extended a scaling theory for flickering buoyant diffusion flames in weakly rotatory flows. It is worth noting that the flame flicker also appears in premixed and partially premixed flames [12, 13].

Dynamical behaviors of multiple flickering flames have recently attracted much research interest. Kitahata et al. [14] first reported that two identical oscillating candle flames exhibit in-phase and anti-phase modes by increasing the distance between the flames. Dange et al.'s flow visualization showed that the interaction between buoyancy-induced vortices plays a significant role in producing different dynamical modes [15]. This vortex-dynamical mechanism was substantiated by numerical simulations [16, 17] and experiments [18, 19] for various (laminar and turbulent) diffusion flames. Okamoto et al. [20] investigated three flickering candle flames in an equilateral triangle arrangement and observed four distinct dynamical modes, such as the in-phase mode, the partial in-phase mode, the rotation mode, and the death mode. Yang et al. [21] computationally reproduced these four modes and interpreted them from the perspective of vortex interaction and particularly of vorticity reconnection and vortex-induced flow. Chi et al. [22] systematically investigated three flickering flames in the isosceles triangle arrangement and identified seven dynamical modes by developing a Wasserstein-space-based methodology for mode recognition. Forrester [23] experimentally observed an initial-arch-bow-initial "worship" oscillation mode for four candles in a square arrangement and conjectured the existence of a chimera state, which is characterized by the hybrids of synchronized and desynchronized flames. For larger systems of flickering flames, there are richer dynamical phenomena, while the flame-vortex and vortex-vortex interactions present bases for causing dynamical flame modes.

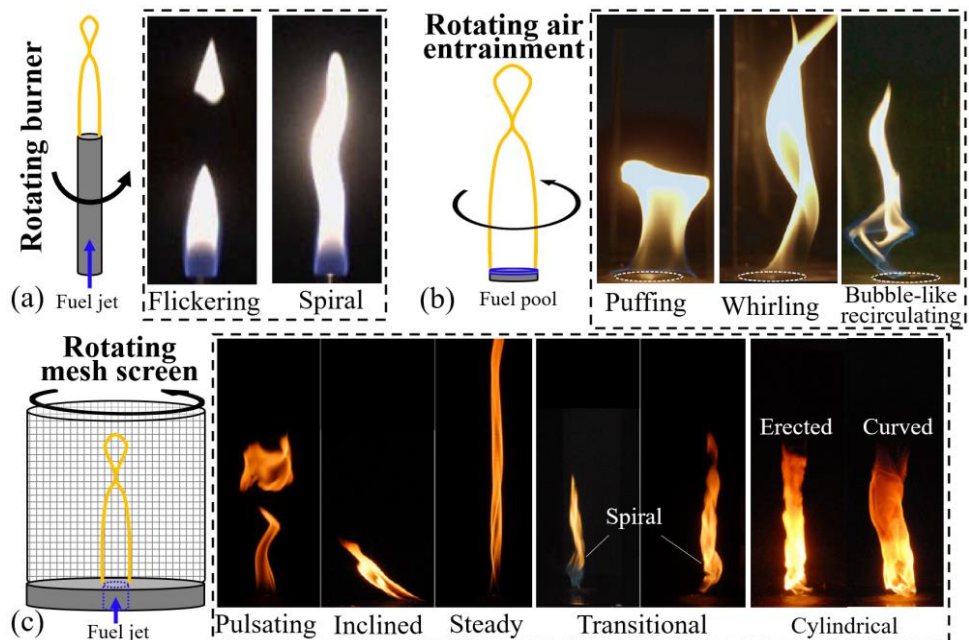

FIG. 1. Various patterns of flickering flames or puffing fires with swirl: (a) the rotating burner [24], (b) the rotating air entrainment due to arranged vanes [25], and (c) the rotating mesh screen [26].

The co-existence of swirling flows and flames is very common in nature and engineering applications. In nature, for example, fire whirls formed by the ambient swirl usually present a powerful but disastrous combustion [27]. In many combustion engines, swirling flows are actively introduced to stabilize the diffusion flames [28]. There is a need to thoroughly understand the impact of swirling flows on flame flicker, leading to many impactful studies both through experimental approaches and numerical analysis. Chuah and Kushida [29] showed that the external vortical flows stabilized the flickering flame and increased the flame height. Gotoda and coworkers [24, 30] studied the influence of burner rotation on flame stabilization and found that the flame flickering could retain at low rotational speeds but transition into low-dimensional deterministic chaos (flames exhibit spiral oscillation, as shown in Fig. 1(a)) at sufficiently large rotation speeds. By generating a spinning flow through the spiral air entrainment, Coenen et al. [25] found that the puffing instability of pool fires could be suppressed in sufficiently strong rotating flows and a helical instability or a vortex bubble might appear instead, as shown in Fig.1(b). Lei et al. [26] utilized the rotation of a mesh screen to generate a vortical flow around a buoyant diffusion flame and identified different flame patterns, such as pulsating, inclined, steady, transitional, and cylindrical flames shown in Fig. 1(c). They observed that the buoyance-driven flame oscillation still existed in the weak fire whirls and the pulsation frequencies were higher than that in the quiescent environment [31].

Despite the above-mentioned noticeable progress toward understanding buoyant diffusion flames subject to externally swirling flow conditions, the dynamical behaviors of flickering buoyant diffusion flames in a large range of swirling flow intensity are still unclear, and the corresponding interpretations based on vortex dynamics were inadequately attempted. Consequently, the present study aims to provide a relatively comprehensive investigation of the dynamical behaviors of flickering buoyant diffusion flames under variable external swirling flow conditions as well as the vortex-dynamical analysis of the different dynamical behaviors. The rest of the paper is organized as follows. In Section 2, the computational methodology for the flickering buoyant diffusion flames and the external swirling flows is expatiated and validated at first. The identified six flame dynamical behaviors and the relevant discussions are presented in Section 3, followed by some concluding remarks in Section 4.

## 2 Computational Methodology

### 2.1 Flickering Buoyant Diffusion Flames

Flickering buoyant diffusion flames are computationally realized and validated first before being imposed in externally swirling flows. As shown in Fig. 2(a), the computational domain is a square column with $16D$ side and $24D$ height, where $D$ is the characteristic length of the gaseous fuel jet. An impermeable, non-slip, and adiabatic solid-wall boundary is used at the bottom ground (grey area), while the central fuel inlet is ejected at the uniform velocity $U_0$. The other sides are set as an open boundary condition, which allows gases to flow into or out of the computational domain depending on the local pressure gradient. To implement a swirling flow, four wind walls are set at the lateral sides, with details provided in the next section. We carried out domain- and mesh-independence studies and adopted a uniform structure mesh of $160 \times 160 \times 240$ for the parametric studies to ensure high accuracy with reasonable computational cost. Our previous study [11, 32] can show that the mesh refinement (each grid has $\Delta \hat{x} = \Delta \hat{y} = \Delta \hat{z} = 10^{-2}$) is sufficient to capture the essential features of the buoyance-induced flicker of a laminar diffusion flame. Particularly, the central region ($8D \times 8D \times 12D$) is further refined in the present cases to analyze closely the flame lift-off behavior due to the emergence of local flame extinction. Fig. S1 shows that the flame frequency of the benchmark case (the flickering buoyant diffusion flame without the local flame extinction) in the $160 \times 160 \times 240$ mesh can be accurately calculated.

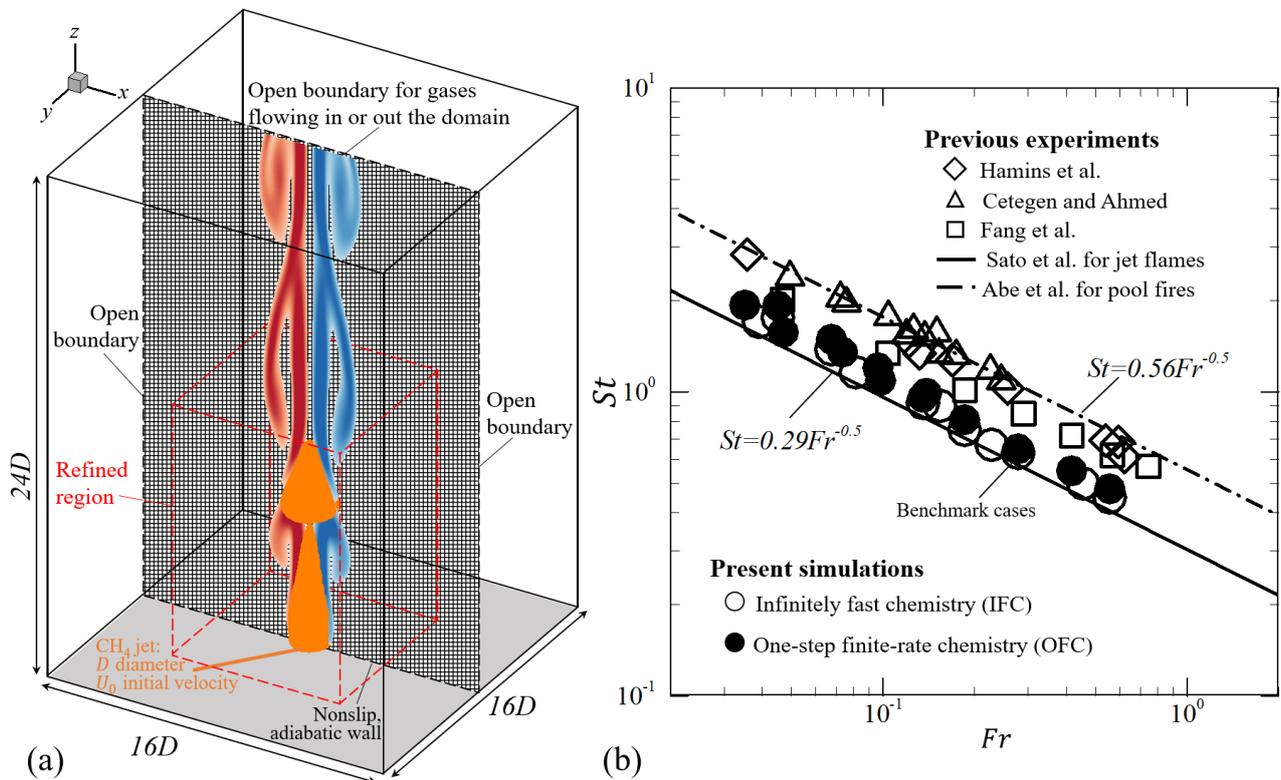

FIG. 2. (a) Schematic of the computational domain, mesh, and boundaries. (b) Validation of single flickering flames with the scaling laws [6, 33] and previous experiments [5, 34, 35]. The numerical results are obtained by using infinitely fast chemistry (IFC) and one-step finite rate chemistry (OFC).

The unsteady, three-dimensional, incompressible (variable-density) flow with chemical heat release was solved using Fire Dynamics Simulator (FDS) [36]. Spatial integration was carried out by using a kinetic-energy-conserving central difference scheme, and the time integration was advanced by an explicit second-order predictor/corrector scheme. Previous computational works [21, 32] have proven the reliability of this computational platform in successfully reproducing a variety of dynamical modes in the dual and triple flickering flame systems.

For pool flames and jet flames with relatively low flow velocities, the flow motion is dominated by buoyance and remains laminar in most of the flow region (it could be turbulent far downstream of the flame). As a result, the flames are far from extinction when the flow characteristic time $\tau_f$ is much larger than the chemical reaction characteristic time $\tau_c$, and the defined Damköhler number of a flickering buoyant diffusion flame can be assumed to be sufficiently large:

$$Da = \tau_f/\tau_c \gg 1 \qquad (2)$$

where the flow characteristic time is defined as the smallest one among various flow time scales such as the buoyance time scale $(D/g)^{1/2}$ and the convection time scale $D/U_0$

$$\tau_f = \min\,[(D/g)^{1/2}, D/U_0] \qquad (3)$$

and the chemical reaction characteristic time $\tau_c$ can be determined by the fastest reaction. Based on the large Damköhler number approximation, the previous computational works [11, 21, 32] were formulated and carried out on the computational platform of FDS. Consequently, the simplified chemistry model (e.g., the mixing-limited chemical reaction model) is hence sufficient for modeling the flickering buoyant diffusion flames with relatively small sizes and reactive fuels, and more sophisticated combustion and turbulence models are usually unnecessary.

In the present problem, the externally swirling flow introduces an additional swirl time scale $1/\Omega$, where $\Omega$ is the angular velocity of the swirling flow. Consequently, Eq. (2) must be updated by

$$\tau_f = \min\,[(D/g)^{1/2}, D/U_0, 1/\Omega] \qquad (4)$$

As a large range of the swirling flow intensity was considered in the present problem, $1/\Omega$ can be so small to invalidate the large Damköhler number assumption, potentially causing local extinction of the flame. To capture the local extinction of diffusion flames and avoid the unnecessary complexity of a detailed reaction mechanism brought to the present problem, however, a prototypical one-step finite-rate reaction model was used:

$$d[F]/dt = -Ae^{\frac{-E_a}{RT_f}}[F]^\alpha[O]^\beta \qquad (5)$$

where $[F]$ ($\alpha$) and $[O]$ ($\beta$) are the concentrations (the reaction orders) of the fuel and the oxidizer respectively. The reaction rate constant is a function of the pre-exponential factor $A$, the activation energy $E_a$, the flame temperature $T_f$, and the universal gas constant $R$. For the convenience of computation, we adopted methane/air combustion [37] as an example, where $A = 1.3 \times 10^9$, $E_a = 202512.4\,J/mol$, $R = 8.314\,J/(K \cdot mol)$, $\alpha = -0.3$, and $\beta = 1.3$. It should be noted that $A$ has a complex dimension in general but it is dimensionless in this example because of $\alpha + \beta = 1$.

The biggest advantage of adopting Eq. (5) is that only one chemical reaction time scale is introduced to the problem, which can be estimated by

$$\tau_c = 1/(Ae^{\frac{-E_a}{RT_f}}) \qquad (6)$$

As a result, the Damköhler number of a flickering buoyant diffusion flame under an externally swirling flow is defined by

$$Da = \frac{\tau_f}{\tau_c} = \min\left[(D/g)^{1/2}, D/U_0, 1/\Omega\right] A e^{\frac{-E_a}{RT_f}} \quad (7)$$

To validate the present computational methodology and models for capturing the flickering phenomenon of buoyant diffusion flames, we conducted many simulations with various $U_0$, $D$, and $g$ in the quiescent environment. In Fig. 2(b), the present results show that the calculated frequencies of flickering flames agree well with previous experiments [5, 12, 34, 35]. The scaling relations of $St = 0.29 Fr^{-0.5}$ and $St = 0.56 Fr^{-0.5}$ are for flickering jet flames [6] and puffing pool fires [33] respectively. In addition, the comparisons between the reaction mechanisms show that the reaction has slight influences on the flickering frequency in the quiescent environment, and this is consistent with that the flame flicker is insensitive to the fuel types and the chemistry [38-40]. Considering that the local extinction tends to occur with increasing intensities of the swirling flow, we used the one-step finite-rate chemistry in all the simulations in the following analysis.

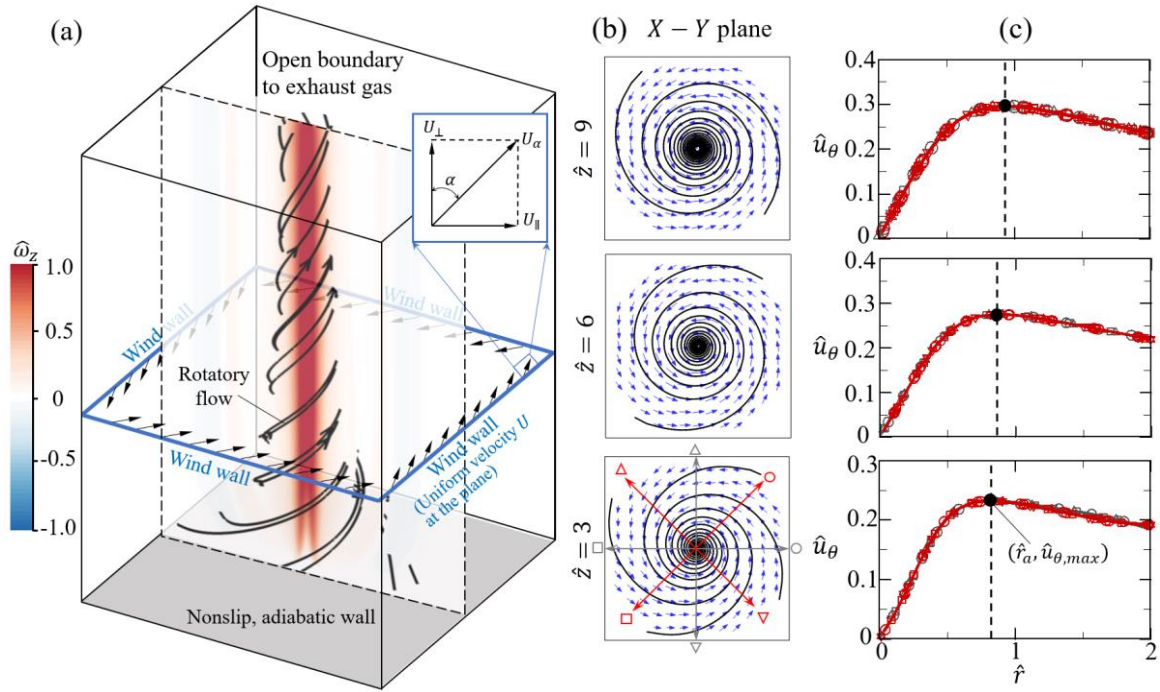

FIG. 3. (a) The rotating flow is imposed from wind walls with inlet velocity $U = (U_\perp, U_\parallel)$, where $\alpha$ is the included angle between velocity components and $R = U/U_0$. The contour of the Y-Z plane shows the vertical component $\widehat{\omega}_z$ of vorticity (b) The flow field (velocity vector and streamline) of the X-Y planes at $\hat{z} = 3$, 6, and 9 for the case at $R = 0.17$ and $\alpha = 45°$. (c) The radial profiles of azimuthal velocity $\hat{u}_\theta$. A vortex core is formed within the radial location $\hat{r}_a$, where increases monotonously up to the maximum $\hat{u}_{\theta,max}$. Eight geometrical symbols denote different azimuth angles distributing uniformly along a 360-degree circle.

## 2.2 Numerical Implementation of Externally Swirling Flows

In the present study, swirling flows are imposed into the computational domain by loading the inlet air on the four lateral sides, namely the "wind wall" as shown in Fig. 3(a). On the wind wall, the inlet velocity is $\boldsymbol{U_\alpha} = \boldsymbol{U_\perp} + \boldsymbol{U_\parallel}$, where $\boldsymbol{U_\perp}$ and $\boldsymbol{U_\parallel}$ are the normal and azimuthal velocity components, respectively, and the angle between $\boldsymbol{U_\parallel}$ and $\boldsymbol{U_\alpha}$ is $\alpha$. Consequently, the airflow circulation is formed in the central region and the central vortical flow can be controlled by adjusting the magnitude $U = |\boldsymbol{U_\alpha}|$ and the angle $\alpha$ of inlet airflow. Similar approaches [41, 42] have been reported to adjust the swirl of the incoming air. To facilitate the following presentation and discussion of results, $D$, $\sqrt{gD}$, and $\rho_\infty$ are used to nondimensionalize all kinematic and dynamic flow quantities. The intensity of the swirling flow is measured by a dimensionless parameter $R = U/U_0$, which is proportional to the nondimensional circumferential circulation of $2\pi \hat{r}_a \hat{u}_{\theta,max}$. The notations of all quantities are consistent with those in the previous study [11].

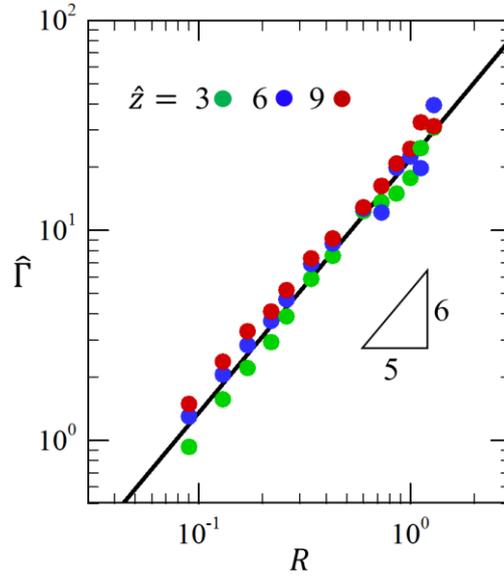

FIG. 4. The correlation between the circulation $\hat{\Gamma}$ and the swirling intension $R$ with the fixed $\alpha = 45°$. The circulation is defined as $\hat{\Gamma} = \int \boldsymbol{u} d\boldsymbol{l} / \sqrt{gD^3}$ along the closed circle $l$ with the radius $\hat{r}_a$ at three cross-sections of $\hat{z} = 3$, 6, and 9.

To validate the swirling flow generated by the present approach, we conducted a series of simulations of non-reacting flows up to $R = 1.30$. As an example, the flow field at $R = 0.17$ and $\alpha = 45$ is shown in Fig. 3(a-b). According to the velocity fields and streamlines in the Y-Z plane and the X-Y planes at $\hat{z} =$ 3, 6, and 9, the vortical flow is characterized by a circular area of significantly concentrated vorticity, forming a vortex core. As shown in Fig. 3(c), the radial profiles of azimuthal velocity $\hat{u}_\theta$ in several X-Y planes are plotted for quantifying the vortical flow fields. Importantly, $\hat{u}_\theta$ linearly increases up to a maximum value (referred to as $\hat{u}_{\theta,max}$) at $\hat{r}_a$ and then gradually decays along the radial direction, which is similar to the experimentally measured transverse flow [43]. For the present vortical flows in a large range of $R$, $\hat{u}_\theta(\hat{r}_a)$ is proportional to $R$ and $\hat{r}_a$ is around 0.8. In addition, Fig. 3 (c) shows the radial profiles of $\hat{u}_\theta$ at eight uniformly distributed azimuth angles are almost the same within the range of $\hat{r} \leq 2$, indicating a good axis symmetry of the flow in the region that is of interest. It is noted that a slight non-symmetry of the flow may occur around the fuel inlet, but this has negligible influence on the flame downstream.

Figure 4 shows that the vortical strengths, defined by the dimensionless circulation $\hat{\Gamma} = \int \mathbf{u} d\mathbf{l}/\sqrt{gD^3}$, are enhanced with increasing $R$ when $\alpha$ is fixed at 45°. In addition, the scaling correlation of $\hat{\Gamma} \sim R^{6/5}$ is valid for a quite wide vertical range up to $\hat{z} = 9$. Figure S2 shows the maximum azimuthal velocity $\hat{u}_{\theta,max}$ and the radial location $\hat{r}_a$ of vortex cores generated at different $R$. It should be noted that the angular velocity of the swirling flow $\hat{\Omega}$ increases with $R$ and is estimated to $\hat{u}_{\theta,max}/(2\pi\hat{r}_a)$.

## 3. Results and Discussion

The present study identified and investigated six dynamical modes of the flickering buoyant diffusion flame in external swirling flows. In this section, we attempted to reproduce flame phenomena observed in the previous experiments [24-26, 31] and answer the following questions: how a flickering buoyant diffusion flame exhibits when it is subject to an externally rotating flow; why does the flame flicker vanish once the rotating intension increases up to a certain degree; what the vortex-dynamical interpretations for the flame variation under different swirling conditions are.

### 3.1 Flickering Flames: Benchmark Cases

Figure 5 shows two benchmark cases for a buoyant diffusion jet flame at $Re = 100$ in a quiescent environment (i.e., no externally swirling flow, $R = 0$), where the infinitely fast chemistry (IFC) assumption is used in Case I, whereas the one-step finite-rate chemistry (OFC) assumption is used in Case II. Due to the instability of flame-induced buoyancy [9, 44], an axis-symmetric toroidal vortex is formed as the growth and roll-up of shearing between the flame sheet and the surrounding air, as shown in Fig. 5(a-b). It should be noted that the flickering flame always keeps axial symmetry during the up-and-down periodic motion, as the streamlines always stay in the plane crossing the central axis. The calculated flickering frequencies are almost the same for the two cases, and the difference is less than 5%. Specifically, these two diffusion flames are nearly identical when comparing their flame behaviors and toroidal vortices during the entire cycle. The formation, growth, and shedding of the vortex correspond to the stretching, necking, and pinch-off of the flame, respectively. It can be seen in Fig. 5(c-d) that the vorticity layer forms at the flame base to stretch the flame (the normalized time $t^*=\hat{t}\hat{f}_0$=0~0.25), curlily grows along and necks the flame ($t^*$=0.25~0.75), and sheds off at $t^*=0.75$ to pinch off the flame. During $t^*=0.75\sim1.0$ the shedding vortex moves downstream and the carried flame bubble burns out soon, while a new vortex generates at the flame base and a new cycle starts. In this way, the flame performs a periodic flickering process.

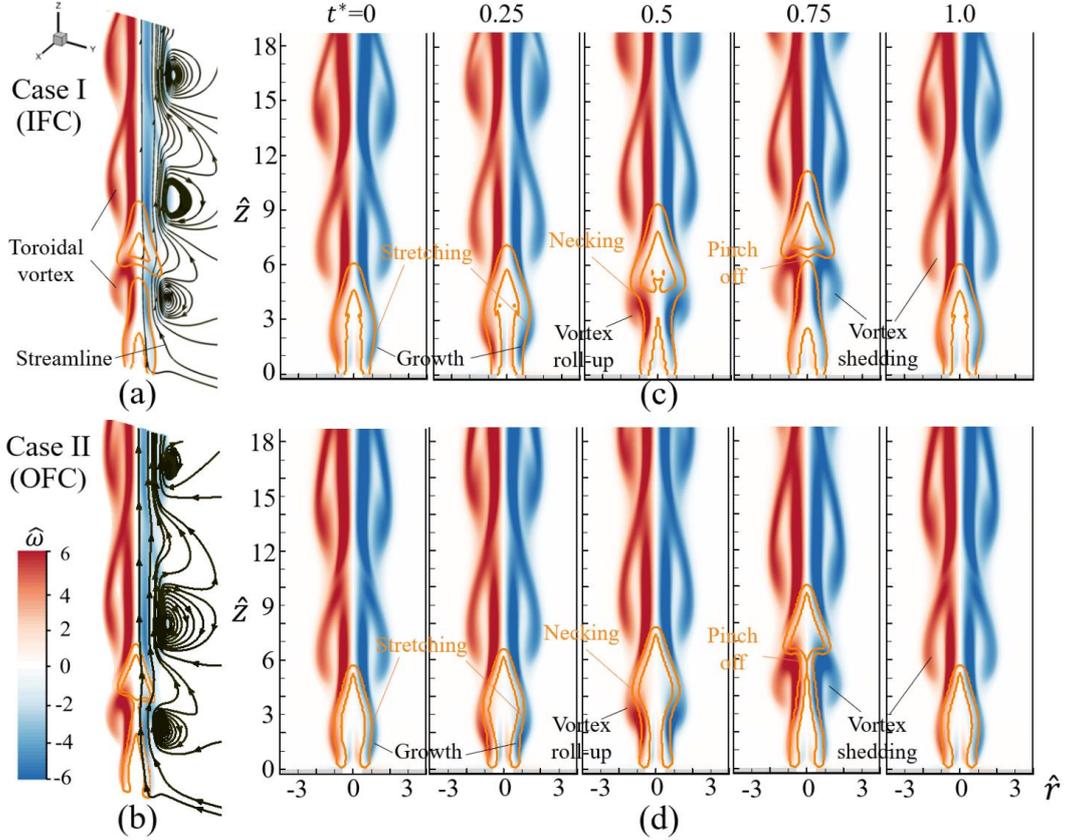

FIG. 5. The contour of vorticity $\hat{\omega}_\theta$ of flickering buoyant diffusion flames for the benchmark cases ($Re=100$, $Fr=0.28$) in Fig. 2: (a) Case I (IFC) and (b) Case II (OFC). The flame is represented by the orange isoline of heat release. The streamlines are plotted around the flame. (c-d) The time-varying evolution of flames and vortices in the benchmark cases.

The phase portraits of flickering buoyant diffusion flames in Case I and Case II are shown in Fig. 6. It is seen that each case has two phase trajectories, which are plotted by the axial components of velocities at three consecutive streamwise locations (i.e., three upstream locations and three downstream locations) in Fig. 6(a-b). During five periodic flickering processes, the upstream and downstream phase portraits for a flickering flame present the same closed ring shape, as shown in Fig. 6(c-f). Most remarkably, the nearly identical portraits in the phase space of Case I and Case II indicate that the chemistry has negligible effects on the dynamical behaviors of flame flicker.

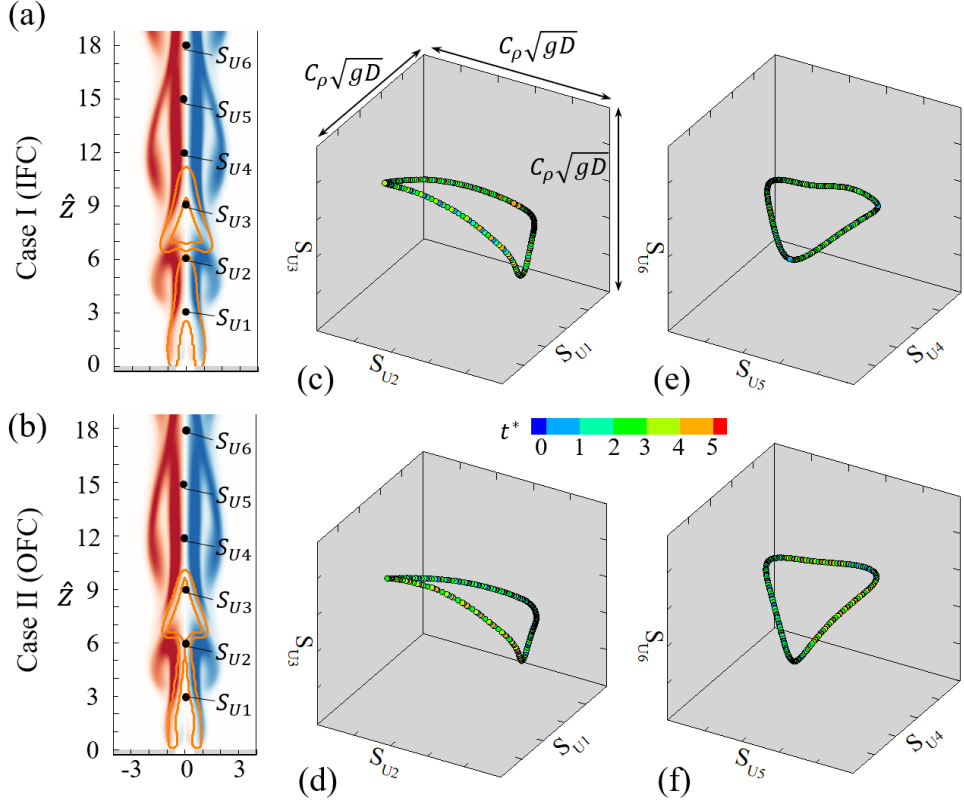

FIG. 6. (a-b) Six axial velocities $S_{U_i}, i =1, 2, 3, 4, 5,$ and 6 at $\hat{z} =3, 6, 9, 12, 15,$ and 18 respectively along the central axis in the benchmark cases. (c-f) Their phase portraits in the cubic space with the same range of $C_\rho\sqrt{gD}$, where $C_\rho = \rho_\infty/\rho_f = 7.5$ is the density ratio of ambient air and flame.

## 3.2 Faster Flickering Flames

Compared with the above benchmark cases of $R = 0.0$, a similar evolution of the toroidal vortex for the case of $R = 0.26$ is shown in Fig. 7, where the stretching, necking, and pinch-off of the flame remain in close synchronization with the formation, growth, and shedding of the vortex. Besides, the phase portraits are still a closed ring shape, which means the external swirling flow does not break the topological structure of the dynamical system.

Interesting observations can be made through a more detailed comparison. First, the flame tends to be pinched off at a further downstream location with increasing $R$. Specifically, the flame pinch-off occurs at $\hat{z} = 6.0$ for the case of $R = 0$ but at $\hat{z} = 6.6$ for the case of $R = 0.26$. Second, the flame pinch-off tends to occur earlier with increasing $R$. Specifically, the flame in the case of $R = 0$ is just pinched off at $t^* = 0.75$, while it has already been pinched off for the case $R = 0.26$. Our previous study [11] addressed this phenomenon in detail and interpreted that the external swirling flow induces an additional vertical flow that expedites the shedding of the toroidal vortex. Third, we plotted streamlines around the flames at $R=0$ (the reference flame) and 0.26 and colored them by the local helicity density $\hat{h} = \hat{\boldsymbol{u}} \cdot \hat{\boldsymbol{\omega}}$. The zero value of $\hat{h}$ means that the streamline is locally orthogonal to the vorticity line, while the nonzero $\hat{h}$ can be used to quantify the local geometrical helix. Specifically, it can be seen in the reference flame that the flow around the flickering flame in a quiescent environment is lamellar because $\hat{h}$ is zero everywhere, and the streamlines are restricted in the plane due to the axis symmetry. As the ambient flow is rotating, the lamellar flow is twisted toward the circumferential direction so that the streamlines tilt out of the plane to form a spiral ring, as shown in Fig. 7(a). Nonzero $\hat{h}$ appears in the vorticity layer around the flame and increases with $R$. Regardless of the local helix of the flow field, the flame morphology retains the approximate axis symmetry to a certain extent.

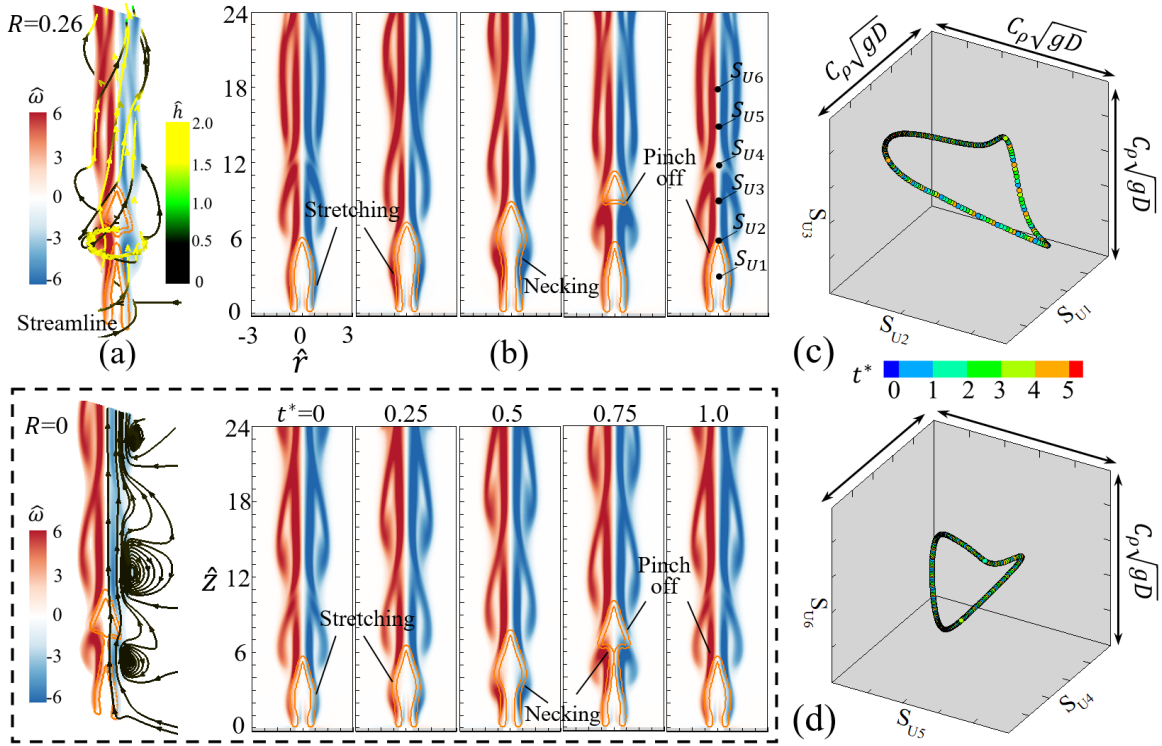

FIG. 7. Faster flicker of the buoyant diffusion flame ($Re=100$, $Fr=0.28$) at $R=0.26$ with the fixed $\alpha=45°$: (a) the flow around the flame, (b) the time-varying evolution of the flame, (c-d) the phase portrait in the cubic space plotted by six velocity components $S_{U_i}$, $i=1, 2, 3, 4, 5$, and 6 at $\hat{z}=3, 6, 9, 12, 15$, and 18 respectively along the central axis. All phase spaces have the same range of $C_\rho\sqrt{gD}$. The benchmark flame at $R=0$ is shown in the dotted rectangle.

As there are negligible changes in calculating the flickering frequency by using the total heat release rate $Q$, or the vertical velocity $u_z$, or the temperature $T$ at a fixed point [11], $Q$ is, without losing the generality, used to determine the frequency of flame flicker. The results show that the periodic oscillation of $Q$ for the case of $R=0$ has a frequency of $f_0=10.1$ Hz, which is smaller than 12.0 Hz for the case of $R=0.26$. The frequency comparison suggests that the external swirling flow causes the flame to be pinched off faster. To reveal the underlying mechanism, Yang and Zhang [11] theoretically modeled the frequency relation for buoyancy-driven diffusion flames in weakly rotatory flows. First, they formulated the generation rate of total circulation $\hat{\Gamma}$ inside a control mass

$$\dot{\hat{\Gamma}} = -[2C_\theta^2 \hat{r}_c \Delta\hat{r} + (C_\rho - 1)\hat{g}\Delta\hat{z}] \tag{8}$$

where $\hat{r}_c$ is the radial position of the vortex layer and close to the flame sheet; the density ratio $C_\rho = \rho_\infty/\rho_f$ is a measurable quantity for a given flame, for example $C_\rho$ is about 7.5 for the present computational methane/air flames, $\Delta\hat{r}$ and $\Delta\hat{z}$ are the radial and vertical unit lengths of the vortex layer associated with the control mass. It should be noted that the first term of Eq. (6) is attributed to the external rotatory flow, which is absent in the theory of Xia and Zhang [9]. Second, they established the periodic formation process of a toroidal vortex associated with the flame flicker

$$\hat{\Gamma}_{TV}^* = C_h Ri\hat{t}^2 + (C_j + C_r)\sqrt{Fr}\hat{t} \tag{9}$$

where $C_h$ is a constant pre-factor for characterizing the buoyancy-induced flow; the constant $C_j$ relies on the configuration and the jet inlet condition; $C_r = 2C_\theta^2 \hat{r}_c \overline{R}/\widehat{U}_0^2$ is a pre-factor for characterizing the external rotational flow. Third, the frequency relation was obtained by applying a threshold for the accumulation of the circulation inside the toroidal vortex (i.e., $\hat{\Gamma}_{TV}^* = C$)

$$\hat{f} = \frac{f}{\sqrt{g/D}} = \frac{1}{2C}\left(C_{jr}Fr + \sqrt{C_{jr}^2 Fr^2 + CC_h C_\rho}\right) \tag{10}$$

where $C_{jr} = C_j + C_r$ is a pre-factor for the combined contributions of the initial jet flow and the external rotatory flow. For the case of $R = 0$, Eq. (10) can degenerate into that of Xia and Zhang. Therefore, the frequency increase of flickering buoyant diffusion flames due to the additional rotation of the ambient flow is expressed as

$$(\hat{f} - \hat{f}_0) \propto R^2 \tag{11}$$

As shown in Fig. 8, the present results show that the increase in flicker frequency obeys the scaling relation of Eq. (11). This finding agrees very well with the scaling theory for flickering buoyant diffusion flames in weakly swirling flows [11]. In physics, the external swirling flow enhances the radial pressure gradient around the flame, and the significant baroclinic effect $\nabla p \times \nabla \rho$ contributes an additional source for the growth of toroidal vortices. Consequently, the toroidal vortices reach the threshold of circulation for shedding early.

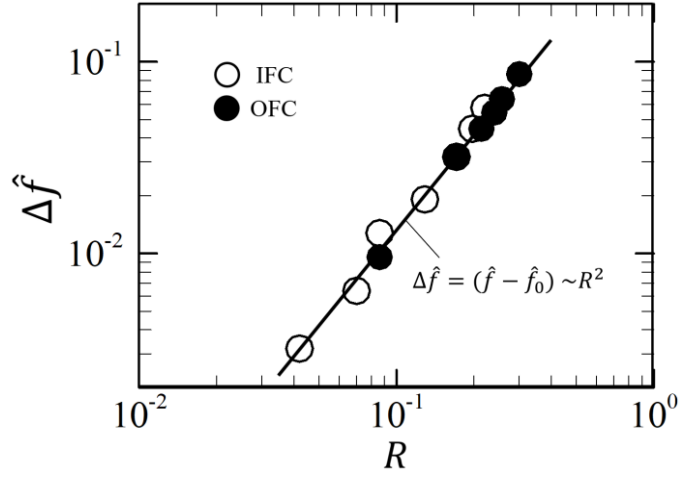

FIG. 8. Comparison between the correlation of $\Delta\hat{f} = (\hat{f} - \hat{f}_0) \sim R^2$ with the numerical results of the infinitely fast chemistry (IFC) and the one-step finite rate chemistry (OFC) in the present study. The swirling flows are fixed at $\alpha = \pi/4$.

## 3.3 Oscillating Flames

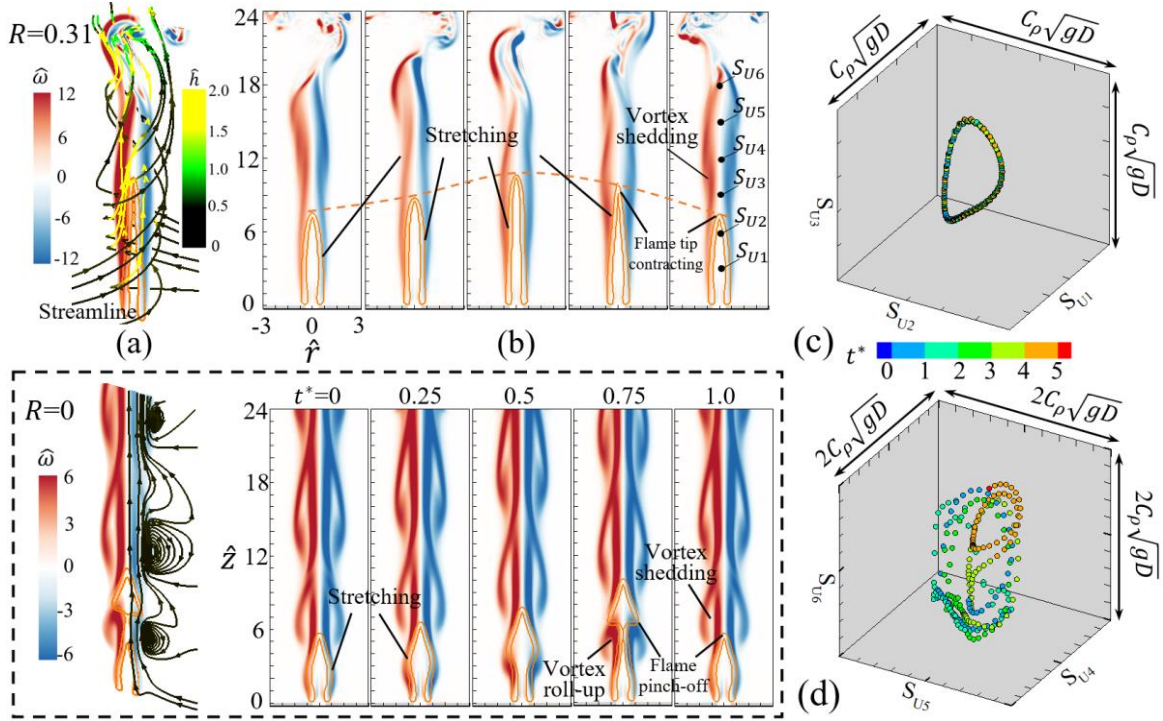

FIG. 9. Tip oscillation of the buoyant diffusion flame ($Re=100$, $Fr=0.28$) at $R = 0.31$: (a) the flow around the flame, (b) the time-varying evolution of the flame, (c-d) the phase portrait in the cubic space plotted by six velocity components $S_{U_i}, i =$1, 2, 3, 4, 5, and 6 at $\hat{z} =$3, 6, 9, 12, 15, and 18 respectively along the central axis. The upstream phase space is twice as large as the downstream is. The benchmark flame at $R = 0$ is shown in the dotted rectangle.

As $R$ increases up to 0.31, the diffusion flame is not pinched off anymore and its top oscillates up and down at a frequency of 13.1 Hz, as shown in Fig. 9. Compared with the benchmark case of $R=0$, the circumferential motion of the ambient air is significant as being indicated by the very high helicity density in flame, and the shear layers are highly stretched so that the roll-up of vortex vanishes. Remarkably, the vortex shedding occurs near the top of the flame. Sato et al. [6, 45] found two similar phenomena of flame flicker without any swirl: one is a "tip flickering" in which no flame separation occurs and the top is merely oscillating or elongating periodically, while the other is a "bulk flickering" where the flame is pinched off at a constant frequency. Specifically, the tip flickering of flame appears under high fuel jet velocity conditions ($Fr \gg 1$), but the bulk flickering corresponds to the low fuel jet velocity ($Fr \ll 1$). This indicates that the external swirling flow has an equivalent influence with the increase of the inlet velocity. In addition, Fig. 9(c-d) shows that phase portraits of oscillating flame vary largely, as the downstream trajectory becomes a distinct disorder in a bigger range while the upstream is a smaller closed ring. The shear layer denoted by the vorticity contour presents that the upstream is compact, but the downstream is unstable and the vortex develops into fragments.

## 3.4 Steady Flames

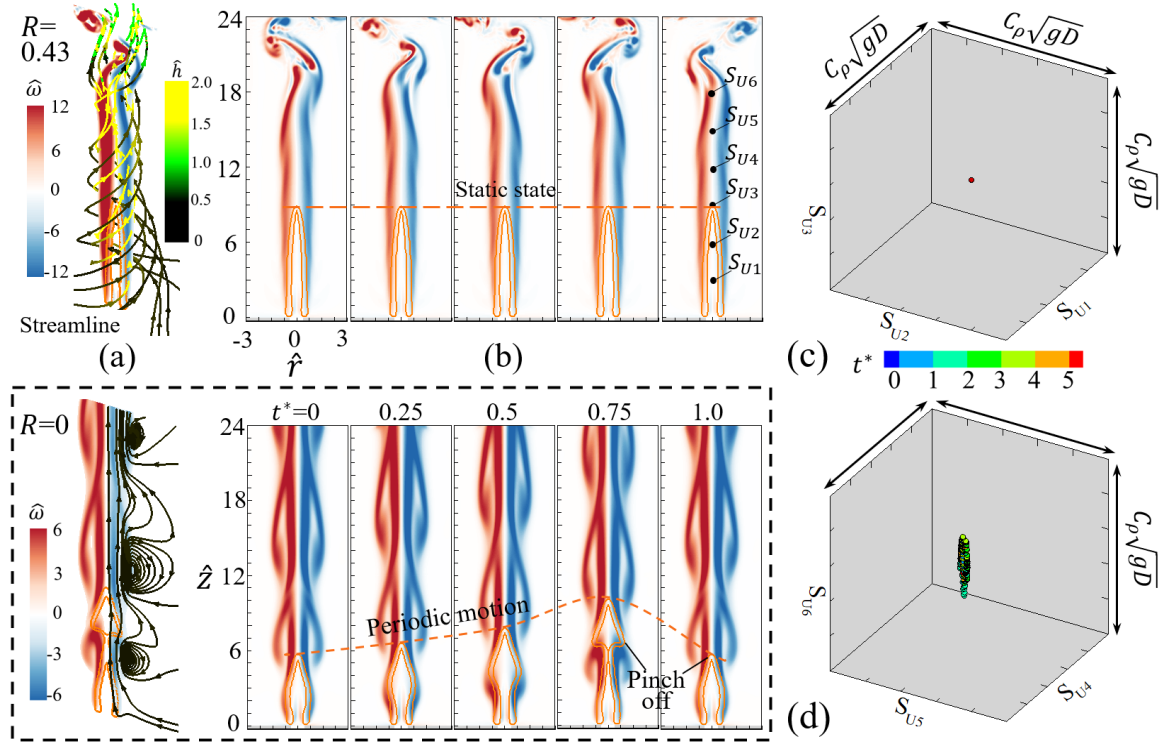

FIG. 10. Steady-state of the buoyant diffusion flame ($Re$=100, $Fr$=0.28) at $R = 0.43$: (a) the flow around the flame, (b) the time-varying evolution of the flame, (c-d) the phase portrait in the cubic space plotted by six velocity components $S_{U_i}, i$ =1, 2, 3, 4, 5, and 6 at $\hat{z}$ =3, 6, 9, 12, 15, and 18 respectively along the central axis. All phase spaces have the same range of $C_\rho \sqrt{gD}$. The benchmark flame at $R = 0$ is shown in the dotted rectangle.

The steady flame at $R = 0.43$ is shown in Fig. 10. Similar to the oscillating flame in the case of $R = 0.31$, the buoyancy-driven flow around the flame is largely twisted compared with that in the benchmark case of $R = 0$. In the highly swirling flow, the flame has no periodic motion, and its shape maintains static all the time, as shown in Fig. 10(b). This observation was also reported in previous studies [26, 29], where the flame flickering can be suppressed by a certain swirling flow. By adding vortical flows around Buk-Schumann diffusion flames, Chuah and Kushida [29] formulated an ideal fire whirl model in that the flame is more stretched and stabilized than the regular diffusion flame. Also, Lei et al. [26] experimentally reported that the weak fire whirl at a small heat release rate is steady with a smooth surface. Accordingly, the upstream phase portrait in Fig. 10(c) degenerates into a point, while the downstream portrait in Fig. 10 (d) shows an oscillation along the $S_{U6}$ direction. It is seen in Fig. 10 (b) that the shear layers are stable for $\hat{z} < 12$ and the instability happens downstream of $\hat{z} = 15$.

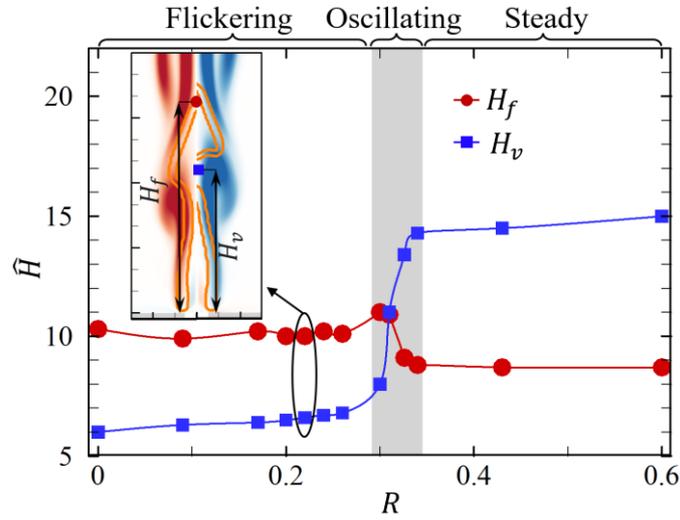

FIG. 11. The vertical position of vortex shedding-off $H_v$ vs the maximum flame height $H_f$ at different $R$.

To facilitate the quantitative comparison between the vortex shedding and the change in flame height, we plot in Fig. 11 the vertical position $H_v$ of the shedding vortex and the maximum flame height $H_f$ at different $R$. The change of the vortex shedding-off has three regimes: within $R < 0.29$, $H_v$ gradually increases with $R$; for $R = 0.29$~$0.35$, $H_v$ increases rapidly; for $R > 0.35$, $H_v$ remains almost constant. Meanwhile, $H_f$ is nearly a constant between 9 and 10, with a slight decrease for $R = 0.29$~$0.35$. These three regions, corresponding to the small, intermediate, and large $R$, correlate with the three flame modes illustrated above: the flickering flame, the oscillating flame, and the steady flame, respectively.

### 3.5 Lifted Flame

As $R$ increases up to 1.11, the diffusion flame does not attach to the bottom due to the local extinction, and the lifted flame is formed with the lifted height of about $4 \sim 6D$. In the present case, the swirl time scale $1/\Omega \sim 10^{-3}s$ is much smaller than the buoyance time scale $(D/g)^{1/2} \sim 10^{-2}s$ and the convection time scale $D/U_0 \sim 10^{-2}s$. Therefore, $Da$ can be estimated to be $A\Omega^{-1}e^{-E_a/(RT_f)} \sim 1$ at $T_f = 1800\ K$. Under the strong swirling flow, the ratio of the residence time to the chemical time becomes very small. During the lift-off process of flame, it can be inferred that there is a balance between reaction and transport phenomena when the circulation reaches a certain value. To further decide the critical circulation of the formation, we need to explore more scenarios where detailed reaction mechanisms and complex flows are considered.

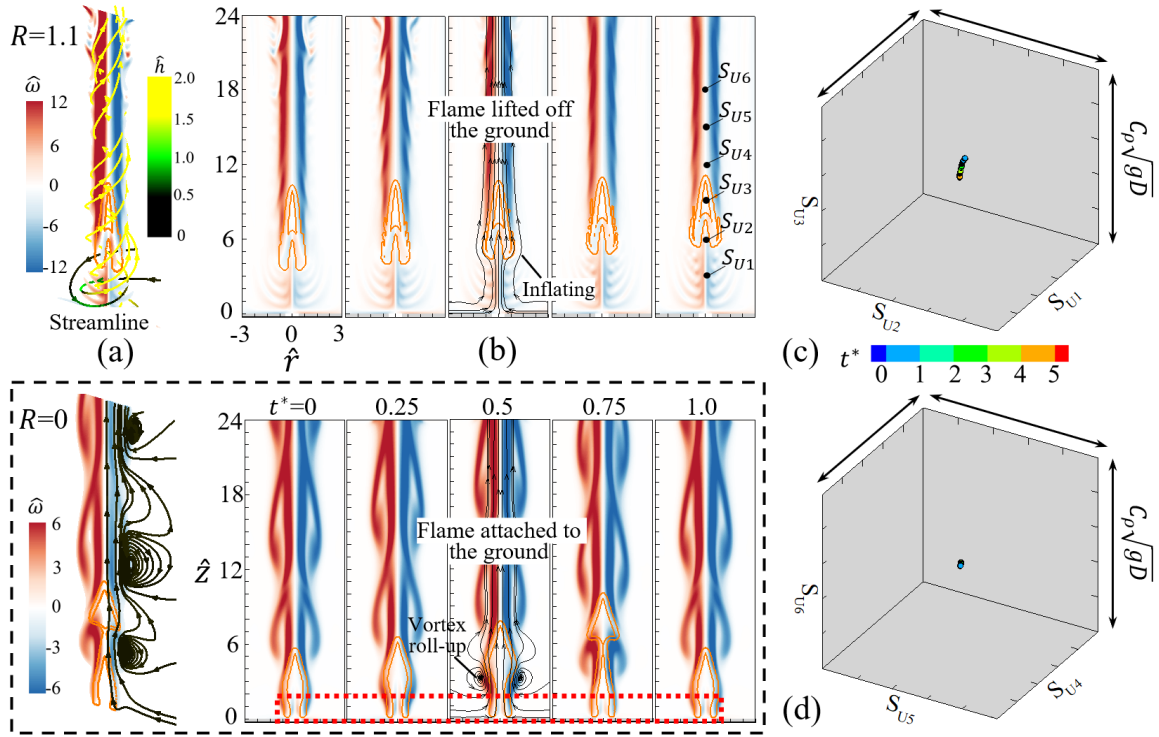

FIG. 12. Lift-off of the buoyant diffusion flame ($Re=100$, $Fr=0.28$) at $R = 1.11$: (a) the flow around the flame, (b) the time-varying evolution of the flame, (c-d) the phase portrait in the cubic space plotted by six velocity components $S_{U_i}, i = 1, 2, 3, 4, 5$, and 6 at $\hat{z} = 3, 6, 9, 12, 15$, and 18 respectively along the central axis. All phase spaces have the same range of $C_\rho\sqrt{gD}$. The benchmark flame at $R = 0$ is shown in the dotted rectangle.

As shown in Fig. 12 (a), the streamlines twine around the flame and have a higher $\hat{h}$ at smaller $\hat{r}$. Besides, the vorticity contour shows that the shear layers intrude into the central region and form a fishbone-like structure. Compared with the benchmark case of $R = 0$, Fig. 12 (b) shows that the flame in the case of $R = 1.11$ is stable and has no oscillation at all. The strongly swirling flow results in a corner-like flow near the fuel inlet, instead of the vortex roll-up induced by the buoyancy in the benchmark flickering flame. The inflating flow with radially outward streamlines is caused by the turning of the incoming flow in the bottom boundary layer. Fig. 12(c-d) shows the nearly motionless phase trajectories of the lifted flame, as the upstream portrait shows a slight change and the downstream one is almost unchanged.

To further understand the formation mechanism of lifted flame, we investigated the flame formation process at the onset of lifted flame. The circulation evolution and the representative flame snapshots during the lift-off formation from the attached flame are shown in Fig. 13, which shows that the circulation first grows and then plateaus around a constant value. From the beginning to $t^*=6$, no change of the circulation is observed, during which the swirling flow is still not imposed upon the central flame. The flame keeps flickering very well ($t_1^*$ and $t_2^*$). From $t^*=6$ to $t^*=11$, the circulation continues to increase gradually, rendering a tall and slender flame ($t_3^*$ and $t_4^*$). After $t^*=11$, the growth of circulation stops, which is accompanied by a steady flame mode ($t_4^*$). During $t^*=11\sim19$, the flame lifts off, the boundary shear layer (a corner flow) intrudes into the central region, and a fishbone-like vortical structure is formed. In the initial stage of lift-off, the flame transitions into a small one ($t_5^*$). Subsequently, the flame stabilizes gradually, and its size becomes big ($t_6^*$).

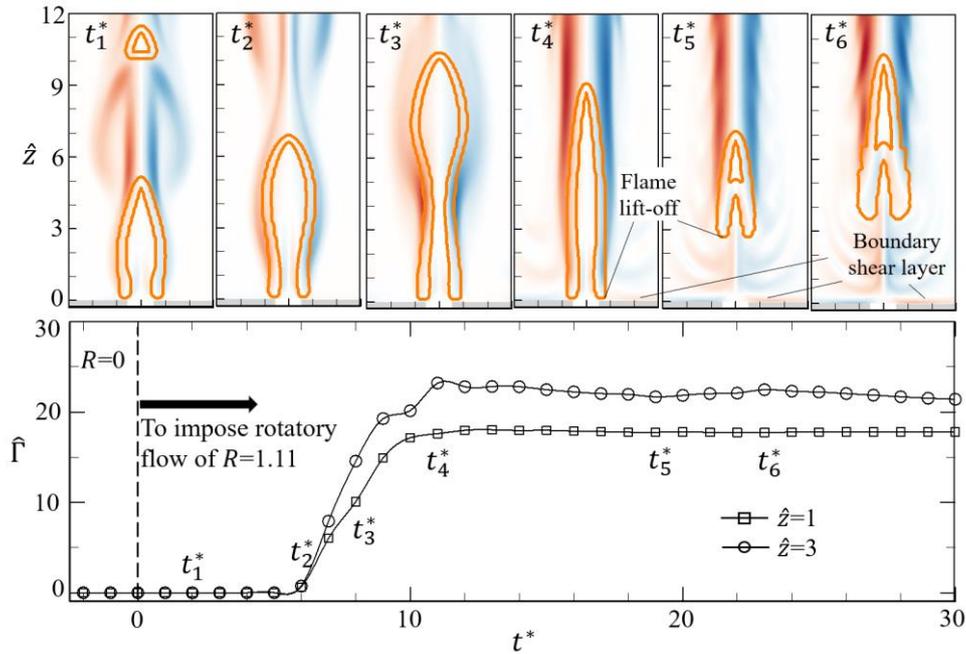

FIG. 13. The variation of circulation with time during the lift-off formation from the attached flame. The circulation is defined as $\hat{\Gamma} = \int \boldsymbol{u} d\boldsymbol{l} /\sqrt{gD^3}$ along the closed circle $\boldsymbol{l}$ with the radius $\hat{r} = 3$. The two lines represent the cross-section at $\hat{z} = 1$ and 3, respectively. Six instantaneous snapshots of flame and vorticity are included.

### 3.6 Spiral Flame

An asymmetric flame was identified for the case of $R = 0.60$ and $\alpha = 79°$, where the flame presents a spiral motion, especially the irregular spin of the top of the flame. In the downstream region, the flame tip inclines apparently, and the nearby streamlines are in the disturbance. As shown in Fig.

14 (b), our simulations captured the dynamical feature of spiral flame. The top of the flame behaves like a swing instead of symmetrically up-and-down for the case of $\alpha = 45°$ and the same $R$. To clearly show the asymmetric swing, the streamlines crossing the flame are plotted within the vorticity contour. In the downstream region of $\hat{z} > 9$, there is a symmetry break of shear layers around the flame, leading to the curved streamlines, while the symmetry of flame, vortical flow, and streamlines retain well for the case of $\alpha = 45°$.

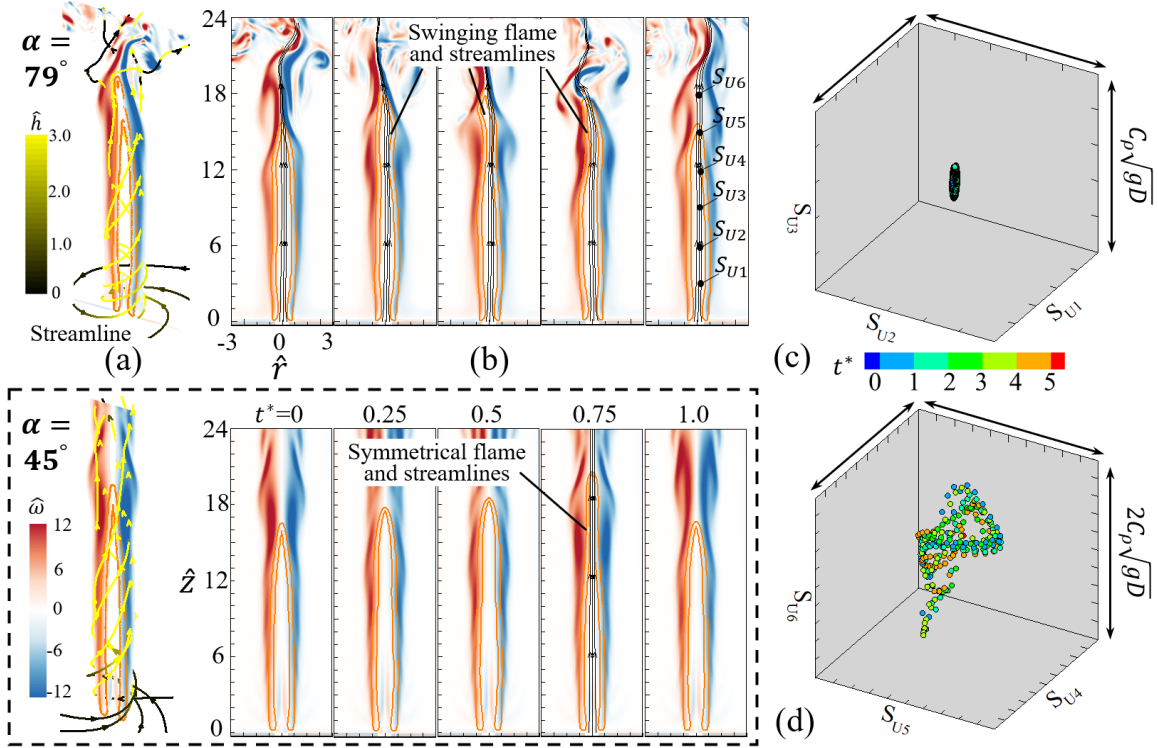

FIG. 14. Spiral structure of the buoyant diffusion flame ($Re = 120$, $Fr=0.40$) in the swirling flow with $R = 0.60$ and $\alpha = 79°$. (a) the flow around the flame, (b) the time-varying evolution of the flame, (c-d) the phase portrait in the cubic space plotted by six velocity components $S_{U_i}, i =1, 2, 3, 4, 5$, and 6 at $\hat{z}$ =3, 6, 9, 12, 15, and 18 respectively along the central axis. All phase spaces have the same range of $C_\rho\sqrt{gD}$. In the dotted rectangle, the oscillating flame ($Re = 120$, $Fr=0.40$) at $R = 0.60$ and $\alpha = 45°$ is shown to facilitate comparison.

Detailed comparisons between the two flames in the three-dimensional view and their orthogonal projections are presented in Fig. 15. The external swirling flows have an impact on the vortical flow due to the formation of the twisted vortex around the flame surface. However, the vortical structure in the spiral flame presents more strong spin than that in the case of $\alpha = 45°$. Particularly, the shedding vortex breaks into some small-scale structures when the swirling flow has a big inlet angle. At the same $|U|$, the bigger $\alpha$ is, the smaller the radial component of the inlet velocity is. Therefore, the central flow in the case of $\alpha = 79°$ is more likely to spread out than that in the case of $\alpha = 45°$. In Fig. 14(c), the upstream phase portrait is a slender ellipse, while the downstream portrait shows a quasi-cycle in the bigger range, particularly $S_{U6}$ at $\hat{z} = 18$. Their two-dimensional projections are shown in Fig. S3. The features of phase portraits are consistent with the dynamic behaviors of the spiral flame.

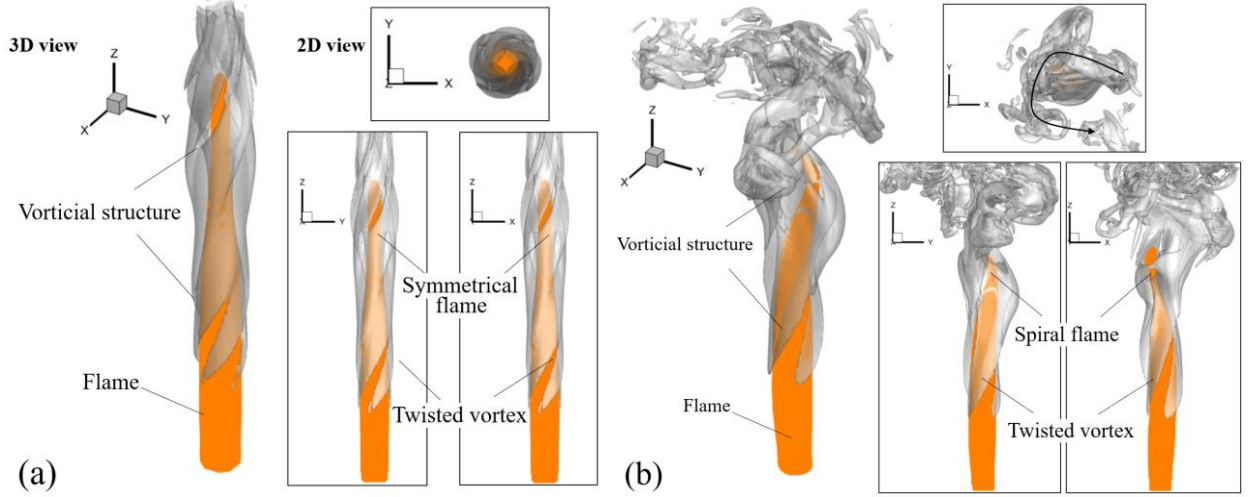

FIG. 15. The three-dimensional view and their three-view drawings of (a) the oscillating flame and (b) the spiral flame corresponding to Fig.14, respectively.

## 3.7 Vortex-bubble Flame

A hat-like flame is captured for the case of $R = 1.30$ and $\alpha = 64°$, in which the lifted flame has a vortex bubble at the flame base. Different from the lifted height $4{\sim}6D$ for the case of $R = 1.30$ and $\alpha = 45°$, the vortex bubble is closer to the fuel inlet and stays close around the vertical position of $2D$, as shown in Fig. 16(a). Particularly, the higher helicity density in the vortex-bubble flame can be observed along the streamlines at the bottom region, compared with that in the case of $R = 1.30$ and $\alpha = 45°$. To clearly illustrate the dynamic features of the vortex bubble flame, the flame and vortex structures within the time interval of $\Delta t^* = 1$ are shown in Fig. 16(b). Different from the lifted flame shown in Fig.12(b), in which the inflating streamlines go through the flame and no circulation exists, the base of the vortex bubble flame kicks outward to form an apparent vortex bubble.

To understand the mechanism underlying the formation of vortex-bubble flame, we noted that the basic definition of vortex breakdown is the abrupt change in vortex structure with retardation along the vortex core and the corresponding divergence of stream surfaces [46, 47]. It is seen from Fig. 16(b) that the streamlines contract in the central region, diverge outwardly, and finally form a convergence toward the downstream. The abrupt change causes a circulation zone where the flow velocities have relatively small magnitudes. In particular, the small vortex cores are generated within the range of $|\hat{u}| < 1$. The bubble region for the vortex breakdown is unstable, and its barycenter and shape vary with time. As shown in Fig. 16(c-d), the phase portraits are a warping string within a relatively small range, instead of a closed topological structure within a relatively big range.

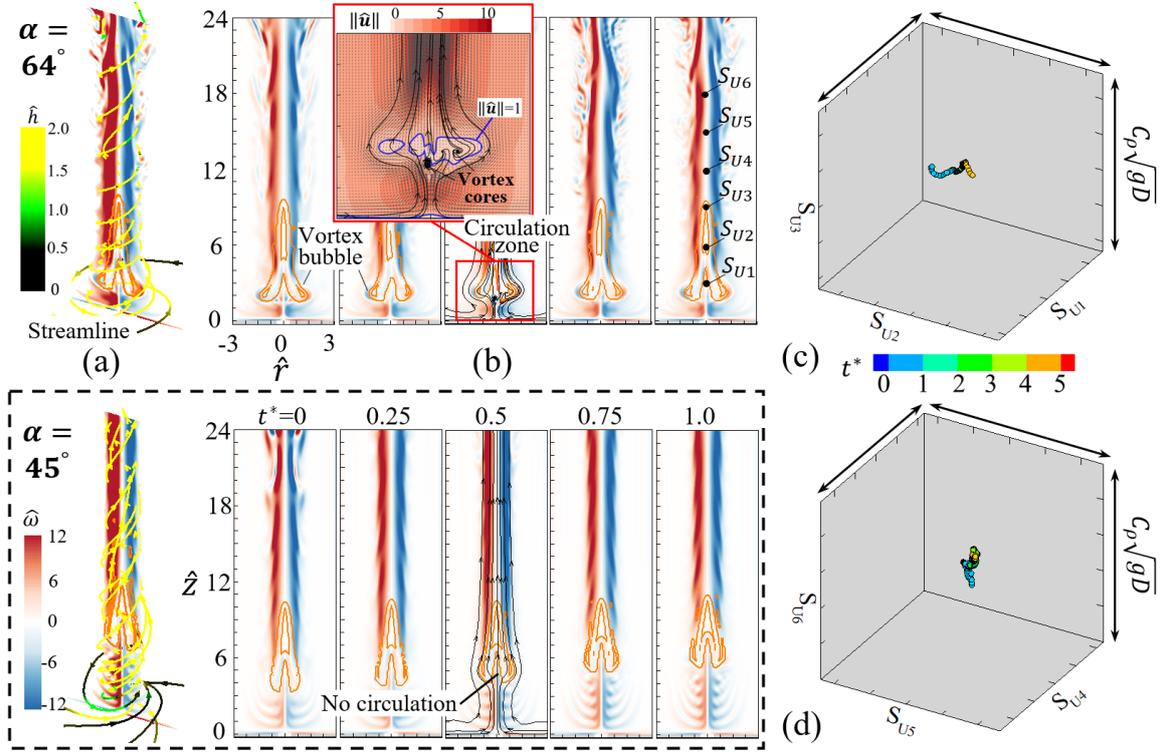

FIG. 16. Vortex bubble of the buoyant diffusion flame ($Re=100$, $Fr=0.28$) in the swirling flow with $R = 1.30$ and $\alpha = 64°$: (a) the flow around the flame, (b) the time-varying evolution of the flame, (c-d) the phase portrait in the cubic space plotted by six velocity components $S_{U_i}$, $i =1, 2, 3, 4, 5,$ and 6 at $\hat{z} =3, 6, 9, 12, 15,$ and 18 respectively along the central axis. All phase spaces have the same range of $C_\rho \sqrt{gD}$. In the dotted rectangle, the lifted flame at $R = 1.30$ and $\alpha = \pi/4$ is shown to facilitate comparison.

As an interesting extension of the present study, Fig. 17 shows a qualitative comparison of the vortex-bubble flame with Chung et al.'s simulation of the blue whirl [48]. Similar features of the flame and the flow are captured in the present simulations. Previous studies [43, 49, 50] suggested that the formation of the blue whirl is accompanied by the occurrence of the vortex breakdown (the bubble, helical, or whirling structures), which is in consistent with the present finding (the flame structure is shaped by the vortex bubble). Besides, the present numerical investigation that the vortex-bubble flame emerges at a relatively large $R$ (it corresponds to the swirling condition of $\hat{\Gamma} > 1$) agrees with the previous experimental observation that the blue whirl falls in a circulation-dominated flow regime. The implications of the vortex-bubble flame for understanding blue whirl merits future studies.

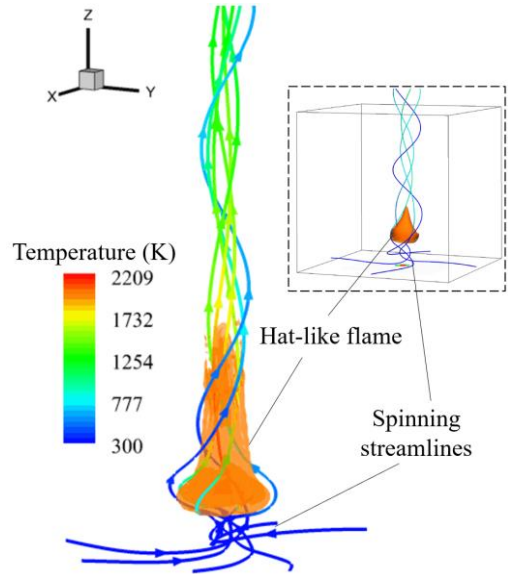

FIG. 17. Comparison of the flame with vortex bubble and the simulated blue whirl (insert figure) [48]. The streamlines are colored by the temperature and the flame is plotted by the iso-surface of the heat release rate at 1 MW/m$^3$.

## 4 Concluding Remarks

This study presents a comprehensive computational investigation of the small-scale flickering buoyant diffusion flames in externally swirling flows with a wide range of swirling intensities. Buoyant diffusion flames exhibit six distinct dynamical modes, namely the flickering flame, the oscillating flame, the steady flame, the lifted flame, the spiral flame, and the flame with a vortex bubble, which were scattered and reported in a few previous experimental studies. The dynamical behaviors of these flame modes are illustrated in the phase space and analyzed from the perspective of vortex dynamics.

In the weak swirl regime, the buoyancy-induced flame flicker becomes faster nonlinearly with increasing $R$, where the increase of $f$ obeys the scaling relation of $(f - f_0) \propto R^2$. Since the vortex shedding occurs the downstream of flame, the flame flicker is suppressed, and instead, the flame has an oscillating tip or remains in a steady state. Higher $R$ can induce local extinction at the flame base and the flame lifts off. This was successfully captured by using a finite-rate chemistry model in the present simulation. In addition, the spiral flame mode and the vortex-bubble flame mode were computationally reproduced by using larger swirl angles.

Based on the understanding that phase portraits are an invaluable tool in studying dynamical systems, different flame modes are illustrated and identified in the phase space. Specifically, the phase portraits for the six modes are summarized as follows: in the flickering flames (the periodic shedding of the toroidal vortex around the flame is the distinct feature), the portraits are the closed ring shape; In the oscillating flames (the toroidal vortex sheds off behind the flame), the upstream portrait is the closed ring, while a big disturbance occurs in the downstream portrait; in the steady flames (no the toroidal vortex), the upstream phase portrait in the steady mode degenerates into a point; In the lifted flames (the shear layers intrude into the central region), the phase portraits are nearly motionless; In the spiral flames (the symmetry break of shear layers emerges), the upstream phase portrait is a small ellipse, while the downstream portrait shows a big quasi-cycle; In the vortex-bubble flames (the vortex breakdown occurs in the flame base), the phase portraits present a warping string.

While the present work provides an understanding of various dynamical modes of diffusion flames, we fully recognize that it does not address the challenging problem of the origin and transition of these modes in a wider parameter space formed by the swirling intensity, the swirling angle, and the Reynolds number. In addition, the influences of turbulence/chemistry interaction in large-scale flames with radiative heat loss were not considered in the present study. The relevant studies merit future work.

## Acknowledgment

This work is supported by the National Natural Science Foundation of China (No. 52176134) and partially by the APRC-CityU New Research Initiatives/Infrastructure Support from Central of City University of Hong Kong (No. 9610601).

## References

[1] D.S. Chamberlin, A. Rose, The flicker of luminous flames, Proc. Combust. Inst. 1-2 (1948) 27-32.
[2] J. Barr, Diffusion flames, Proc. Combust. Inst. 4 (1953) 765-771.
[3] E. Zukoski, B. Cetegen, T. Kubota, Visible structure of buoyant diffusion flames, Proc. Combust. Inst. 20 (1985) 361-366.
[4] J. Buckmaster, N. Peters, The infinite candle and its stability—a paradigm for flickering diffusion flames, Proc. Combust. Inst. 21 (1988) 1829-1836.
[5] B.M. Cetegen, T.A. Ahmed, Experiments on the periodic instability of buoyant plumes and pool fires, Combust. Flame 93 (1993) 157-184.
[6] H. Sato, K. Amagai, M. Arai, Diffusion flames and their flickering motions related with Froude numbers under various gravity levels, Combust. Flame 123 (2000) 107-118.
[7] D. Moreno-Boza, W. Coenen, A. Sevilla, J. Carpio, A. Sánchez, A. Liñán, Diffusion-flame flickering as a hydrodynamic global mode, J. Fluid Mech. 798 (2016) 997-1014.
[8] L.-D. Chen, J. Seaba, W. Roquemore, L. Goss, Buoyant diffusion flames, Proc. Combust. Inst. 22 (1989) 677-684.
[9] X. Xia, P. Zhang, A vortex-dynamical scaling theory for flickering buoyant diffusion flames, J. Fluid Mech. 855 (2018) 1156-1169.
[10] M. Gharib, E. Rambod, K. Shariff, A universal time scale for vortex ring formation, J. Fluid Mech. 360 (1998) 121-140.
[11] T. Yang, P. Zhang, Flickering Buoyant Diffusion Flames in Weakly Rotatory Flows, arXiv preprint arXiv:2208.09278, (2022).

[12] D. Durox, T. Yuan, F. Baillot, J. Most, Premixed and diffusion flames in a centrifuge, Combust. Flame 102 (1995) 501-511.
[13] N. Fujisawa, T. Okuda, Effects of co-flow and equivalence ratio on flickering in partially premixed flame, Int. J. Heat Mass Transfer 121 (2018) 1089-1098.
[14] H. Kitahata, J. Taguchi, M. Nagayama, T. Sakurai, Y. Ikura, A. Osa, Y. Sumino, M. Tanaka, E. Yokoyama, H. Miike, Oscillation and synchronization in the combustion of candles, J. Phys. Chem. A 113 (2009) 8164-8168.
[15] S. Dange, S.A. Pawar, K. Manoj, R. Sujith, Role of buoyancy-driven vortices in inducing different modes of coupled behaviour in candle-flame oscillators, AIP Adv. 9 (2019) 015119.
[16] A. Bunkwang, T. Matsuoka, Y. Nakamura, Similarity of dynamic behavior of buoyant single and twin jet-flame (s), J. Therm. Sci. Technol. 15 (2020) 1-14.
[17] T. Tokami, M. Toyoda, T. Miyano, I.T. Tokuda, H. Gotoda, Effect of gravity on synchronization of two coupled buoyancy-induced turbulent flames, Phys. Rev. E 104 (2021) 024218.
[18] N. Fujisawa, K. Imaizumi, T. Yamagata, Synchronization of dual diffusion flame in co-flow, Exp. Therm Fluid Sci. 110 (2020) 109924.
[19] L. Changchun, L. Xinlei, G. Hong, D. Jun, Z. Shasha, W. Xueyao, C. Fangming, On the influence of distance between two jets on flickering diffusion flames, Combust. Flame 201 (2019) 23-30.
[20] K. Okamoto, A. Kijima, Y. Umeno, H. Shima, Synchronization in flickering of three-coupled candle flames, Sci. Rep. 6 (2016) 1-10.
[21] T. Yang, Y. Chi, P. Zhang, Vortex interaction in triple flickering buoyant diffusion flames, Proc. Combust. Inst., (2022) (in press).
[22] Y. Chi, T. Yang, P. Zhang, Dynamical mode recognition of triple flickering buoyant diffusion flames in Wasserstein space, Combust. Flame 248 (2023) 112526.
[23] D.M. Forrester, Arrays of coupled chemical oscillators, Sci. Rep. 5 (2015) 1-7.
[24] H. Gotoda, T. Ueda, I.G. Shepherd, R.K. Cheng, Flame flickering frequency on a rotating Bunsen burner, Chem. Eng. Sci. 62 (2007) 1753-1759.
[25] W. Coenen, E.J. Kolb, A.L. Sánchez, F.A. Williams, Observed dependence of characteristics of liquid-pool fires on swirl magnitude, Combust. Flame 205 (2019) 1-6.
[26] J. Lei, N. Liu, Y. Jiao, S. Zhang, Experimental investigation on flame patterns of buoyant diffusion flame in a large range of imposed circulations, Proc. Combust. Inst. 36 (2017) 3149-3156.
[27] A. Tohidi, M.J. Gollner, H. Xiao, Fire whirls, Annu. Rev. Fluid Mech. 50 (2018) 187-213.
[28] S. Candel, D. Durox, T. Schuller, J.-F. Bourgouin, J.P. Moeck, Dynamics of swirling flames, Annu. Rev. Fluid Mech. 46 (2014) 147-173.
[29] K.H. Chuah, G. Kushida, The prediction of flame heights and flame shapes of small fire whirls, Proc. Combust. Inst. 31 (2007) 2599-2606.
[30] H. Gotoda, Y. Asano, K.H. Chuah, G. Kushida, Nonlinear analysis on dynamic behavior of buoyancy-induced flame oscillation under swirling flow, Int. J. Heat Mass Transfer 52 (2009) 5423-5432.
[31] J. Lei, N. Liu, K. Satoh, Buoyant pool fires under imposed circulations before the formation of fire whirls, Proc. Combust. Inst. 35 (2015) 2503-2510.
[32] T. Yang, X. Xia, P. Zhang, Vortex-dynamical interpretation of anti-phase and in-phase flickering of dual buoyant diffusion flames, Phys. Rev. Fluids 4 (2019) 053202.
[33] H. Abe, A. Ito, H. Torikai, Effect of gravity on puffing phenomenon of liquid pool fires, Proc. Combust. Inst. 35 (2015) 2581-2587.
[34] A. Hamins, J. Yang, T. Kashiwagi, An experimental investigation of the pulsation frequency of flames, Proc. Combust. Inst. 24 (1992) 1695-1702.
[35] J. Fang, J.-w. Wang, J.-f. Guan, Y.-m. Zhang, J.-j. Wang, Momentum-and buoyancy-driven laminar methane diffusion flame shapes and radiation characteristics at sub-atmospheric pressures, Fuel 163 (2016) 295-303.
[36] K. McGrattan, S. Hostikka, R. McDermott, J. Floyd, C. Weinschenk, K. Overholt, Fire dynamics

simulator user's guide, NIST Special Publication 1019 (2013) 1-339.
[37] C.K. Westbrook, F.L. Dryer, Chemical kinetic modeling of hydrocarbon combustion, Progress in energy and combustion science 10 (1984) 1-57.
[38] K. Sahu, A. Kundu, R. Ganguly, A. Datta, Effects of fuel type and equivalence ratios on the flickering of triple flames, Combust. Flame 156 (2009) 484-493.
[39] H.G. Darabkhani, J. Bassi, H. Huang, Y. Zhang, Fuel effects on diffusion flames at elevated pressures, Fuel 88 (2009) 264-271.
[40] M. Bahadori, L. Zhou, D. Stocker, U. Hegde, Functional dependence of flame flicker on gravitational level, AIAA journal 39 (2001) 1404-1406.
[41] J.B. Mullen, T. Maxworthy, A laboratory model of dust devil vortices, Dynamics of Atmospheres and Oceans 1 (1977) 181-214.
[42] E.J. Kolb, An Experimental Approach to the Blue Whirl, University of California, San Diego2018.
[43] Y. Yang, H. Zhang, X. Xia, P. Zhang, F. Qi, An experimental study of the blue whirl onset, Proc. Combust. Inst., (2022) (in press).
[44] H. Zhang, X. Xia, Y. Gao, Instability transition of a jet diffusion flame in quiescent environment, Proc. Combust. Inst. 38 (2021) 4971-4978.
[45] H. Sato, K. Amagai, M. Arai, Flickering frequencies of diffusion flames observed under various gravity fields, Proc. Combust. Inst. 28 (2000) 1981-1987.
[46] M. Hall, Vortex breakdown, Annu. Rev. Fluid Mech. 4 (1972) 195-218.
[47] O. Lucca-Negro, T. O'doherty, Vortex breakdown: a review, Progress in energy and combustion science 27 (2001) 431-481.
[48] J.D. Chung, X. Zhang, C.R. Kaplan, E.S. Oran, The structure of the blue whirl revealed, Sci. Adv. 6 (2020) eaba0827.
[49] H. Xiao, M.J. Gollner, E.S. Oran, From fire whirls to blue whirls and combustion with reduced pollution, Proceedings of the National Academy of Sciences 113 (2016) 9457-9462.
[50] Y. Hu, S.B. Hariharan, H. Qi, M.J. Gollner, E.S. Oran, Conditions for formation of the blue whirl, Combust. Flame 205 (2019) 147-153.